# Structural characterization of as-grown and quasi-free standing graphene layers on SiC

R. A. Bueno[1], I. Palacio[1], C. Munuera[1], L. Aballe[2], M. Foerster[2], W. Strupinski[3], M. García-Hernández[1], J. A. Martín-Gago[1], M. F. López*[1]

[1] *Materials Science Factory, Instituto de Ciencia de Materiales de Madrid (ICMM-CSIC), Sor Juana Inés de la Cruz 3, E-28049 Madrid, Spain*

[2] *ALBA Synchrotron Light Facility, Carrer de la llum 2-26, Cerdanyola del Vallès, Barcelona 08290, Spain*

[3]*Faculty of Physics, Warsaw University of Technology, Koszykowa 75, 00-662 Warsaw, Poland*

* mflopez@icmm.csic.es

## Abstract

We report on a comparative structural characterization of two types of high quality epitaxial graphene layers grown by CVD on 4H-SiC(0001). The layers under study are a single layer graphene on top of a buffer layer and a quasi-free-standing graphene obtained by intercalation of hydrogen underneath the buffer layer. We determine the morphology and structure of both layers by different complementary in-situ and ex-situ surface techniques. We found the existence of large islands in both samples but with different size distribution. Photoemission electron microscopy (PEEM) measurements were performed to get information about the chemical environment of the different regions. The study reveals that monolayer graphene prevails in most of the surface terraces, while a bilayer and trilayer graphene presence is observed at the steps, stripes along steps and islands.

*Keywords: graphene, SiC, LEEM/PEEM, STM, CVD, H-intercalation*

# 1. INTRODUCTION

Because of its extraordinary physical properties, graphene has emerged as one of the most promising candidates for the development of electronics devices in the new carbon-era beyond the Si CMOS technology. In particular, it combines remarkable electronic properties with a 2D configuration compatible with lithography techniques well developed and used in the field of electronics. Since its discovery in 2004, graphene has been grown on many different substrates [1]. However, one of the most interesting systems for producing electronic carbon devices is the epitaxially grown graphene layers on silicon carbide (SiC) [2-5].

Few layer graphene (FLG) on SiC can be grown by various techniques like thermal decomposition reaction in ultrahigh vacuum (UHV), annealing in an induction furnace under low vacuum or atmospheric pressure conditions, additional carbon supply similar to molecular beam epitaxy, and by molecular beam epitaxy [6-12]. Epitaxial graphene has also been grown on SiC surfaces by chemical vapor deposition (CVD) [13]. The CVD method produces a high quality monolayer graphene on top of the buffer layer, where the $(6\sqrt{3}\times6\sqrt{3})R30º$ superperiodicity of the buffer layer is observed. The graphene produced by this method presents large crystalline atomic terraces, it is much less sensitive to SiC surface defects (higher electron mobility) and allows a better control of the number of graphene layers during growth on both the Si-face and the C-face [13]. This is a strong advantage, because the electronic properties of FLG depend on the thickness. Nevertheless, the interface layer also named $0^{th}$ layer or buffer layer, influences the epitaxial graphene growth on SiC(0001). The graphene becomes n-doped, i.e. the Fermi level is shifted upwards with the Π-bands shifted into the valance band [14].

In order to avoid the graphene-substrate interaction, that quenches partially the graphene properties, a process of hydrogen intercalation can be performed. This leads to the formation of a quasi-free-standing monolayer graphene (QFSMG). Hydrogen intercalates under the interface layer where it reacts with the dangling bonds of the silicon atoms and, as a result, passivates the SiC surface layer. As a consequence, the interface or buffer layer is decoupled from the substrate forming a QFSMG on top [15]. The hydrogenation of graphene is a reversible process since the intercalated hydrogen atoms can be desorbed at temperatures above 900ºC. Interestingly, this easy method for fabricating large scale graphene wafers, produces a QFSMG which, in contrast to epitaxial graphene, is slightly p-doped. These results suggest the possibility of tailoring the electronic properties of graphene on SiC substrates, which will open promising possibilities for nanoelectronic applications.

Local nanoscale studies of the surface and interface structure of CVD epitaxial graphene on SiC were previously performed using several techniques as Raman, X-ray photoelectron spectroscopy (XPS), kelvin probe force microscopy measurements (KPFM), secondary ion mass spectrometry and Hall efect [13,15-20]. Those systematic studies on structural characterization on both graphene as-grown and hydrogen intercalated obtained by patented CVD method on SiC substrates were performed for the first time, enabling optimization of the growth for targeting electronic applications. However, although these techniques can be used in parallel to provide understanding of the chemical and structural properties of graphene, they generally lack the spatial resolution.

On the other hand, to study the micro and nanoscopic structural characteristics of the CVD graphene layer on SiC is of crucial importance because the attributes at the micro and nanoscale will have direct consequences on the properties of the final system. Thus, in order to optimize the large scale growth of CVD graphene, a deep knowledge of the structure at the nanoscale is of paramount importance. For these reasons, in this work we have performed a comparative structural characterization of two types of graphene samples grown by CVD on 4H-SiC(0001), namely single layer graphene (SLG) and QFSMG. To determine the differences in topography and morphology we have used several techniques, such as low-energy electron microscopy (LEEM), low-energy electron diffraction (LEED), LEEM-IV, scanning tunnelling microscopy (STM) and atomic force microscopy (AFM). These techniques give information on the surface characteristics at different length and surface magnitudes from the nano to the micro scales. Additionally, some insights into the chemical environment and electronic properties of both systems have been obtained by means of photoemission energy electron microscopy (PEEM), XPS and KPFM.

## 2. EXPERIMENTAL DETAILS

In this study we have investigated two different kinds of epitaxial graphene samples grown on 4H-SiC(0001) [21]. For both samples, a buffer layer was grown by CVD on 4H-SiC(0001) at 1600°C under an argon (Ar) laminar flow in a hot-wall Aixtron VP508 reactor. The CVD growth method relies critically on the creation of the flow conditions in the reactor that control the Si sublimation rate, adjusting parameters that inhibits Si escape, and enable mass transport of hydrocarbon to the SiC substrate through the argon gas boundary layer. Thus, for the CVD growth, carbon atoms were provided externally by methane gas delivered to the reactor by an argon carrier gas and deposited (with epitaxy) on the SiC substrate [13]. The method, which is different from Si sublimation, offers the precision of synthesizing a pre-defined number of carbon layers, including a single layer on the Si-face of SiC, and is less sensitive to SiC surface defects than the sublimation method [13,15]. From one side, for the SLG samples the previous procedure was carried on until achieving a graphene layer on top of the buffer one. On the other hand, the QFSMG samples were obtained by intercalating H under the buffer layer. This latter protocol transforms the buffer layer into graphene and yields Hall mobility values above 8000 $cm^2$/V·s suggesting that it could be used in electronic applications.

To perform the experiments, the samples were introduced into a UHV chamber with a base pressure lower than $5 \times 10^{-10}$ mbar. Prior to measurements, each sample was cleaned in the UHV equipment to remove the contamination. The cleaning procedure consisted in annealing the sample at 350ºC during 15 minutes. This annealing treatment was carried out by electron bombardment heating. The sample temperature was estimated by thermocouple positioned in the sample holder and confirmed by using an IR pyrometer (emissivity of 0.9). The cleanliness of the sample was checked by the quality of the corresponding LEED patterns.

For the STM measurements, a commercial room temperature STM (Omicron) and chemically etched W tips have been used. The LEEM and PEEM experiments were carried out at the CIRCE beamline of the ALBA Synchrotron, Barcelona (Spain). By using an Elmitec LEEM/PEEM III microscope with electron imaging energy analyzer, lateral spatial resolution close to 20 nm with X-ray excited photoelectrons and 10 nm in LEEM and UV-PEEM modes. XPS spectra with submicron spatial resolution and electron energy resolution ~ 0.2 eV can be obtained [22]. For a quantitative analysis,

XPS C-1s curves were fitted using standard Gaussian-Lorentzian lines and the corresponding integral background. For reproducing the asymmetric line shape of $sp^2$ carbon a Doniach and Sunjic line shape with an asymmetry parameter of 0.068 was used [23].

A commercial Atomic Force Microscopy (AFM) system, from CSI Instrument, operating in ambient conditions was employed to perform morphological and surface potential (KPFM) characterization of the samples. Measurements have been acquired in dynamic mode, using the amplitude as the feedback channel for topography acquisition. Surface potential maps have been measured using single pass KFM mode (HD-KFM) with Au-coated tip from Mikromasch (HQ:NSC18/Cr-Au). Image analysis was performed with the WSxM free software [24].

## 3. RESULTS

Figure 1 shows STM images and LEED patterns obtained for the two kinds of samples. Both samples are optimal for STM characterization as the surface roughness is extremely low and the number of monoatomic steps and surface defects is very small. Figures 1a and 1b show high-resolution STM images of epitaxial SLG grown on a clean 4H-SiC(0001) sample. Atomic resolution is achieved in both images. The large scan image (Fig. 1a) displays an enhancement of the intensity following a quasi-(6x6) periodicity related to the presence of the buffer layer, together with a small ripple corresponding to carbon atoms. Fig. 1b shows a higher magnification image, where the graphene lattice can be more evident. The 6x6 protrusions observed in the STM image are detected all over the sample, indicating that the graphene layer has a buffer layer underneath, and that indeed we have a SLG. All these structural features can be recognized in the LEED pattern of Fig. 1c, where the pink arrow corresponds to the graphene spots, the blue arrow points to SiC spots less intense than the graphene ones and rotated 30º, and finally the green arrow indicates the spots corresponding to the (6√3x6√3)R30º periodicity.

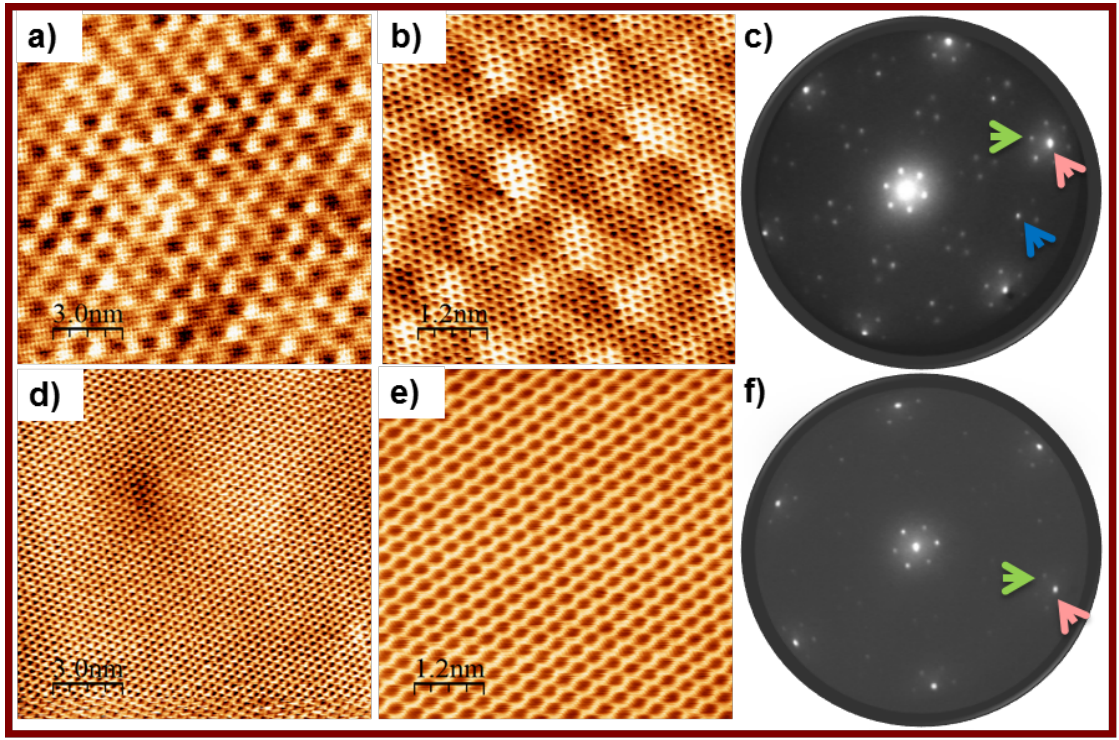

**Figure 1.** *(a) STM image of SLG, scanned area 15 nm x 15 nm, -283mV, 1.51 nA and (b) close view of the previous surface (6nm x 6nm) with atomic resolution, -510.9 mV, 0.36 nA, (c) the corresponding LEED pattern at 42 V. (d) STM images of QFSMG, scanned area 15 nm x 15 nm, -294 mV, 1.366 nA and (e) close view with atomic resolution of the previous surface (6nm x 6nm), -243 mV, 1.33 nA. (f) the corresponding LEED pattern at 42 V.*

Figs. 1d and 1e correspond to the STM images of the QFSMG sample. In this case, as it was already mentioned, a H layer is intercalated between the graphene and the SiC, removing the dangling bonds that give place to the quasi (6x6) reconstruction seen in the previous Figs. 1a-b. Figure 1e shows an atomic resolution image recorded in a smaller area, where only the graphene honeycomb structure of QFSMG is observed, without any trace of the (6x6) periodicity, demonstrating that in the hydrogenation process the buffer layer has been quenched by the H atoms. The LEED pattern of the QFSMG sample is shown in Fig. 1f. A comparison with the LEED of the SLG shown in Fig. 1c reveals a decrease in the intensity of some spots and in some cases their vanishing. Thus, we can observe for the QFSMG a strong intensity at the spots corresponding to graphene (pink arrow), whereas faint spots belonging to buffer layer are shown, and also, the SiC pattern almost disappears. All these effects indicate the successful decoupling of the buffer layer.

After the STM results, we can conclude that at an atomic scale level both kinds of samples exhibit a high degree of perfection. For a characterization at the nanometer scale XPEEM/LEEM images were recorded. LEEM is suitable for analysing these samples since graphene thickness contrast is observed as a function of the electron energy. In fact, the number of graphene layers directly correlates with the number of dips in the electron reflectivity spectra. Additionally XPEEM will provide information about the chemical nature of the different regions.

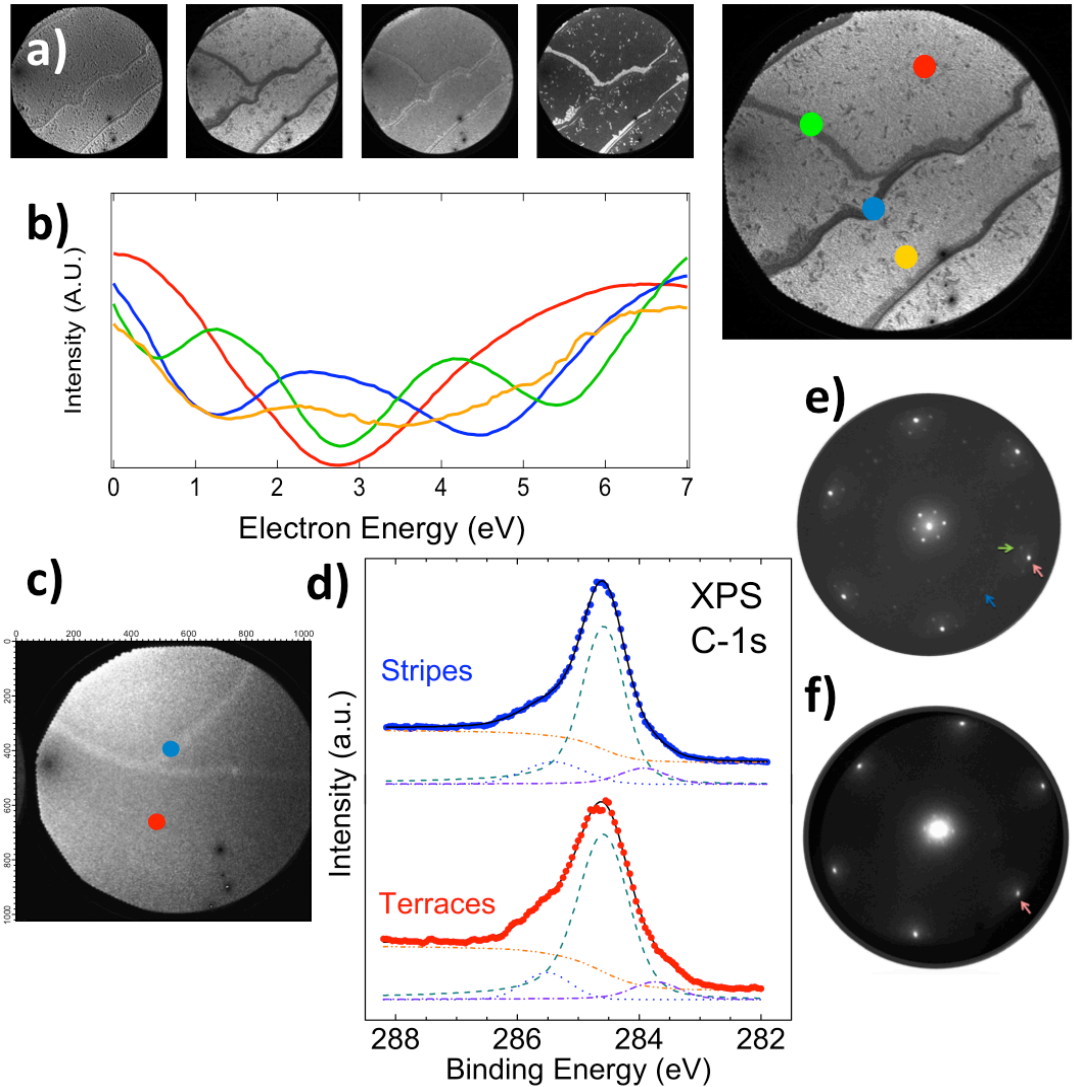

**Figure 2.** *(a) LEEM images of the same area for the SLG sample at a field of view (FOV) of 20µm, taken at different electron energies. In the last image of the sequence four regions of the surface are highlighted: terrace in red colour, island in yellow colour, dark stripe in blue colour and light stripe in green colour. (b) LEEM-IV reflectivity curves on the regions marked by the colour dots in last LEEM image. (c) XPS-PEEM image taken at the C-1s peak energy of the blue spectra in (d). (d) C-1s*

*XPS peak of two different domains: terrace (red line) and stripe (blue line). The spectra were obtained by integrating all equivalent points. µ-LEED patterns at 42 eV corresponding to the (e) terrace area and (f) stripe area, illumination aperture size for incoming electron beam 5µm. The main points are indicated with arrows.*

Figure 2a shows a sequence of LEEM images obtained on the same area of the SLG sample at different electron energies, 0.46, 1.29, 2.13, 2.96 and 5.4 eV, from left to right. The measurement of these images at a sequence of energies was performed with the aim of distinguishing the structural properties of different areas of the surface, since it is known that with this technique the different structural phases have different electron reflectivity as a function of energy [25]. Indeed, in the last image of the series (larger image) four different structures can clearly be distinguished. The most extended one, about 73% of the surface, consists of flat terraces (marked with a red dot), which are 20-50 µm long and around 5 µm wide. Inside the terraces appear small islands approximately 0.5 µm wide (marked yellow), which correspond to the second region and cover only a 3% of the surface. Terraces are limited by steps, where there are two kinds of stripes forming the other two additional types if regions. One of them, seen as light stripes of about 1 µm wide in the figure, covers 13% of the sample (marked green). The other one, seen as dark stripes of about 0.5 µm wide, covers the 11% of the sample (marked blue).

From LEEM images sequentially obtained while the electron beam energy was changed, we observe that at low energy (0.46 eV) the reflectivity of terraces and stripes is very similar (light grey) while the small islands stand out in dark colour. When we increase the energy (1.29 eV), the reflectivity of terraces and islands begins to homogenize while stripes are highlighted. At intermediate energies (2.13 eV), terraces and islands have just completely homogenized and stripes almost disappear. At higher electron energies (2.96 eV) the behaviour of the two kinds of stripes is different, one is indistinguishable from the dark terraces while the other acquires a light colour. Finally, as we already mentioned, at high energies (5.04 eV) all the different structures are detectable. Indeed, the terraces appear with light grey colour, the small islands inside terraces are visible with a darker shade and the two kinds of stripes are clearly differentiated.

Moreover, LEEM is a very useful technique to determinate the number of graphene layers on SiC and discerning this number along the surface sample by taking advantage of its spatial resolution. The method consists in identifying the number of graphene layers by the number of minima from in the electron reflectivity spectra [26,27]. In the case of graphene on SiC the number of minima corresponds to the number of conduction graphene layers being identified as mono- (ML, corresponds to 1 minimun), bi- (BL, corresponds to 2 minima), tri-layer graphene (TL, corresponds to 3 minima), etc. In Figure 2b the LEEM-IV reflectivity curve exhibits periodic oscillations for the different regions of the sample. The data indicate that large terraces (red spectrum) have one minimum in the reflectivity curve (ML), the small islands (yellow spectrum) as well as the dark stripes (blue spectrum) have two minima (BL) and finally the light stripes (green spectrum) have three minima (TL). This result suggests the predominance of ML graphene terraces with the presence of BL graphene islands and stripes and finally some stripes or bands formed by TL graphene. Interestingly, the LEEM data reveal the presence of small islands of BL within the ML terraces. Although BL and TL graphene have been detected frequently in the steps and stripes regions for epitaxial graphene grown by Si sublimation, it is not frequent to observe BL graphene within the ML terraces. A possible explanation of the presence of this BL islands in the terraces could rely on the graphene growth method. The CVD method of epitaxial graphene growth offers a much higher precision of synthesizing a pre-defined number of carbon layers, including a single layer on the Si-face of SiC. With CVD, the nucleation sites for graphene growth are located at the atomic steps, therefore enabling step-flow epitaxy. The accurate optimization of growth time and hydrocarbon precursor partial pressure allows to minimize the presence of BL areas on the surface. In case of Si sublimation, the non-uniformity of graphene thickness is too high to observe such behavior. Nevertheless, these small bilayer inclusions do not deteriorate the carrier mobility which for CVD graphene on SiC was achieved at the level of 6000 -7000 cm$^2$/Vs at RT, as a typical value for large scale samples [28].

In order to understand the chemical nature of the stripes, we performed X-ray photoemission electron microscopy (XPEEM). In XPEEM, the sample is illuminated by X-rays of a given energy, and a photoemission map is recorded with the emitted photoelectrons. In our case, the photoelectrons corresponding to the C 1s core level peak were selected to obtain the photoemission elemental map. Figure 2c shows an XPEEM image collected at the C1s core level of the SLG sample. In this map, we can distinguish two areas: terraces (red circle region) and stripes (blue circle region). The stripe regions appear brighter than the terrace ones indicating a more intense C1s signal.

This is in agreement with the reflectivity results, where TL and BL graphene, with higher C contents, were detected for these areas. The XPS spectra of C1s collected in both regions are represented in Figure 2d after proper normalization. The solid black line through the data points as well as the subspectra are the results of the least-squares fit. For both cases, stripes and terraces regions, three components were used to simulate the spectra, corresponding to the different chemical environment of the carbon atoms. The main subspectra around located around 284.5 corresponds to the C sp$^2$ emission from the graphene layer, a small shoulder around 283.7 eV is related to the C atoms in the SiC bulk crystal and a shoulder around 285.4 eV is assigned to C atoms in the buffer layer. Comparing with previous XPS works on a similar system [18], our data exhibit a larger signal for the graphene layer with respect to that of the bulk SiC. This effect could be expected since the XPEEM works with electron kinetic energies (58-70 eV in this case) which have a shorter mean free path than those of the previous XPS measurements given by electron kinetic energies around 1200 eV.

The LEEM-PEEM instrument allows us to measure µ-LEED by confining the incoming electron beam by an aperture on a small area of the sample, acquiring LEED patterns in selected micrometric regions. Figure 2e shows the µ-LEED pattern recorded at the terraces area. The image exhibits bright spots typical of graphene (marked with a pink arrow), weak spots corresponding to SiC grid (marked with a blue arrow) and to the (6√3x6√3)R30º reconstruction (marked with a green arrow). This pattern agrees with that of ML graphene on a SiC substrate. On the other hand, Figure 2f shows the µ-LEED pattern recorded in an area with high stripe contents revealing intense spots of the graphene hexagon pattern, while spots belonging to SiC and to the reconstruction have disappeared. This pattern is the one expected for a region with several graphene layers, confirming thus the assignment from the LEEM reflectivity data.

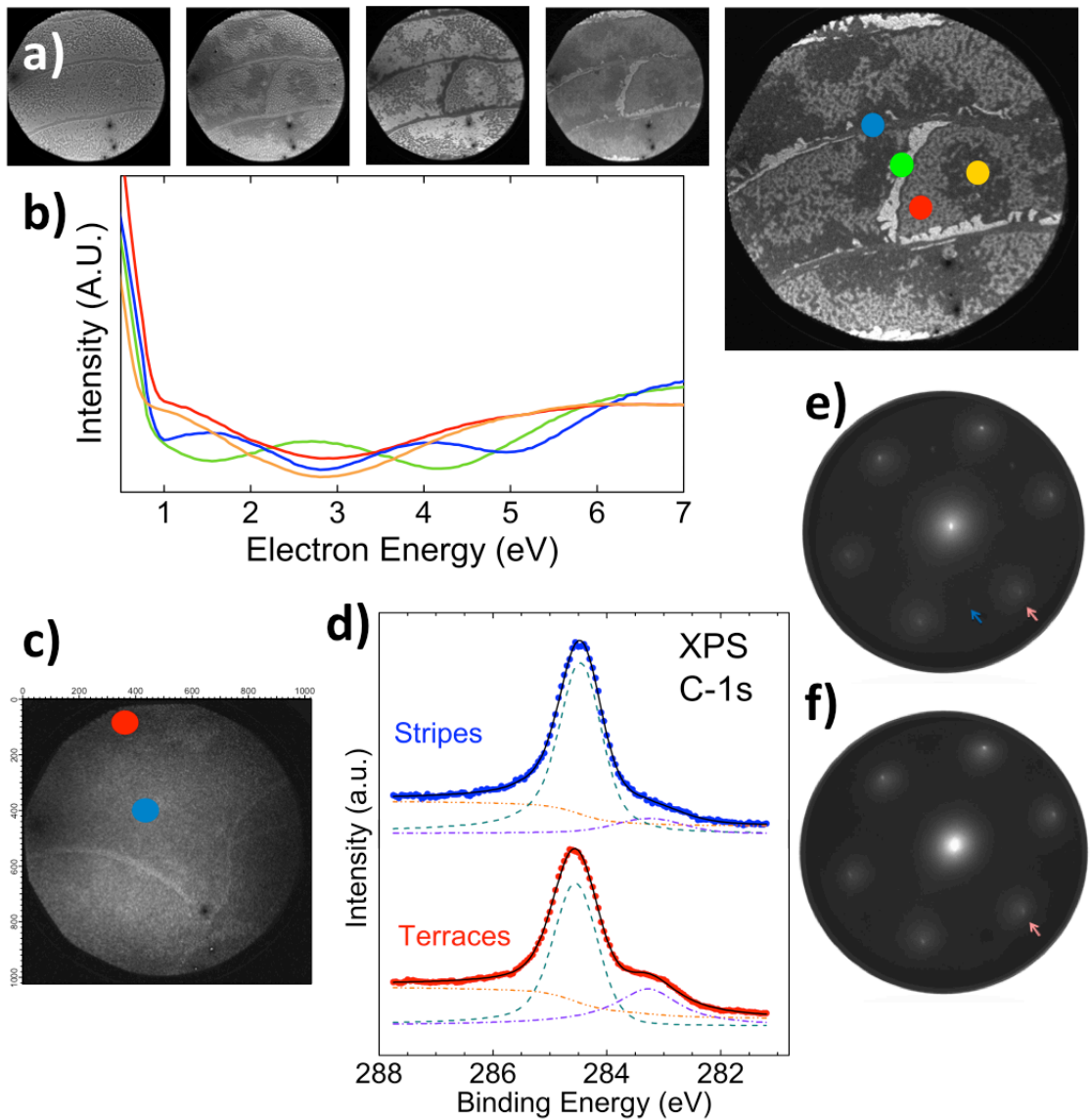

**Figure 3**. *(a) LEEM images of the QFSMG sample in the same area, FOV=20μm, taken at different electron energies. In the last image of the sequence four types of regions are highlighted: terrace in red colour, patches in yellow colour, dark stripe in blue colour and light stripe in green colour. (b) LEEM-IV reflectivity curves on the regions marked by the colour dots in last LEEM image. (c) PEEM image taken at the C-1s peak energy of the spectra in (d). (d) C-1s XPS spectra of two different domains: terrace (marked red) and stripe (marked blue). μ-LEED patterns (5 μm diameter areas) at 62 eV corresponding to the (e) terrace and the (f) stripe areas. The main spots are highlight with arrows.*

An analogous characterization was carried out for the hydrogenated sample to allow for a proper comparison. The LEEM and PEEM characterization of the hydrogenated sample, the QFSMG sample, is shown in Figure 3. The reflectivity behaviour of the QFSMG sample when the electron energy is varied (0.8, 2.8, 4.8, 9.2, 11.6 eV from left to right) can be observed in the Figure 3a series. At low energy (0.8 eV) the reflectivity of the sample is quite homogeneous. By increasing the energy (2.8 eV) some patches stand out within the terraces in dark grey. At higher energies (4.8 eV), four different areas can be distinguished: terraces (dark grey) with patches (light grey), steps with near stripes (dark grey) and far stripes (light grey). At 9.2 eV the contrasts are exchanged, stripes and terraces are bright while the patches and previously light stripes become darker. Finally, at 11.6 eV, the contrast has fully reversed and we can perfectly distinguish the four zones again. In particular, analysing this last and large image, we determine that most of the sample, about 87% of the surface, is composed of terraces (marked in red), which are 40-50 μm long and about 15 μm wide, with some patches inside (marked in yellow) with ≈ 5 μm of extension. These patches occupy a 44% of the sample surface, while the region of the terraces free of patches covers a 43% of the surface. Like in the previous samples, the terraces are limited by steps and in the vicinity of those steps two kinds of stripes appear. From one side, light stripes cover the 9% of the sample (marked in green) and are about 1 μm wide. On the other, dark stripes cover the 4% of the sample (marked in blue) and are about 0.5 μm wide. The main difference with respect to the previous type of samples, as depicted from the LEEM sequence images of Fig. 3a, appears in the terraces, where two types of regions with different reflectivity are distinguished.

The graphene layer thickness was also characterized for this sample using LEEM-IV, as the number of graphene layers can be determined from the reflectivity oscillations. Such LEEM-IV reflectivity curves of the different domains for the QFSMG sample are shown in Figure 3(b). As both the terraces (red spectrum) and the patches (yellow spectrum) have one minimum in the reflectivity data (ML), the evident contrast in reflectivity of both regions must be ascribed to a heterogeneous content of intercalated hydrogen. Dark stripes (blue spectrum) have three minima (TL) and light stripes (green spectrum) have two minima (BL).

PEEM image of Figure 3(c) shows mainly two areas for the QFSMG sample: terraces (red circle) and stripes (blue circle). A brighter signal for the case of the stripes region can be observed, suggesting a carbon accumulation near the steps forming stripes. This result is in agreement with the LEEM-IV reflectivity data where the stripes near the steps exhibited bi- and tri-layer graphene. The corresponding XPS spectra of the C1s core levels are represented in Figure 3(d) where the solid black line through the data points as well as the subspectra are the result of the least-squares fit. Differently to Figure 2(d), for both regions, only two components were needed to simulate the curves. The main feature, similar to the SLG sample, is located around 284.5 eV and corresponds to the C sp$^2$ signal of the graphene layer. On the other hand, no presence of emission corresponding to the C of the buffer layer is observed, as it is expected for this case where the buffer layer has been transformed into a graphene layer due to the hydrogen intercalation. Finally, the subspectra at around 283.3 eV is related to the SiC bulk emission. A comparison of the SiC bulk emission for both samples, SLG and QFSMG, points to a shift towards lower binding energies for the hydrogenated sample. As already reported [29] this shift is due to the band bending taking place between the substrate and the graphene layer, related to the doping of QFSMG samples on H-terminated SiC samples. However, in our case the shift is lower than in previous works, probably due to the fact that the hydrogen intercalation process has not been completely homogeneous.

The μ-LEED patterns recorded at the terraces and at the stripes areas are shown in Figures 3(e) and 3(f), respectively. The terraces pattern exhibits bright spots corresponding to the graphene (pink arrow), together with faint spots corresponding to SiC (blue arrow), while the spots corresponding to the (6√3x6√3)R30º reconstruction are not observed. This result is in agreement with the lack of a buffer layer due to the H intercalation. On the other hand, the μ-LEED stripes pattern shows only the spots of graphene (pink arrow). This is a consequence of having less graphene layers in this region, as in the SLG sample.

AFM measurements were also performed in the different regions of the two types of samples, combining topographic and surface potential imaging (KPFM mode). The latter imaging mode has become a very useful tool for showing nanoscale variations in the graphene thickness homogeneity [30,31]. Figure 4 presents simultaneously acquired topography and surface potential maps for both, SLG and QFSMG samples. Both samples present similar morphology, showing the characteristic features of graphene grown on the Si-face of SiC, which is strongly dominated by SiC terraces separated by step bunches, with heights ranging from 4 nm to 10nm as shown in Figure 4(a) and (b). Clusters with 5-10 nm height covering the QFSMG surface are due to ambient contamination of the samples, likely adsorbed due to the higher number of islands and defects with respect to the SLG. Simultaneously acquired surface potential maps are presented in Figure 4(c) and 4(d). For both samples, brighter areas exhibiting higher surface potential values are distinguished, that directly correlate with the terrace edges of the corresponding topography image. By the LEEM analysis presented before it has been confirmed that bilayer/trilayer patches grow along these terrace edges. The surface potential distribution along the terraces also shows a direct correlation with the LEEM images, with a dark background and islands presenting a brighter contrast. The measured surface potential difference between terrace and edges is $40 \pm 10$ mV for both samples. This correlation between surface potential contrast (from KPFM measurement) and number of layers (from LEEM analysis) has been ascribed to different substrate induced doping levels, as a result of the different energy dispersions of mono- and bilayer graphene [20,32,33]. Additionally, for the case of SLG a possible origin of the different doping of graphene at the steps regions could be a different interface structure between graphene and the substrate, with a local delamination of the buffer layer on the steps [34,35]. Finally, it has to be also considered the influence of the different response to environmental doping between single and bilayer graphene [36,37]. The inhomogeneous surface potential measured at the terraces of the QFSMG sample might also originate from a heterogeneous content of intercalated hydrogen content, as pointed out above. As has been studied previously, the intercalated hydrogen may be responsible of a change in the doping character of the graphene layer [15]. Figure 4(d) in fact points to an inhomogeneous doping of the graphene layer in the QFSMG sample due to hydrogen intercalation.

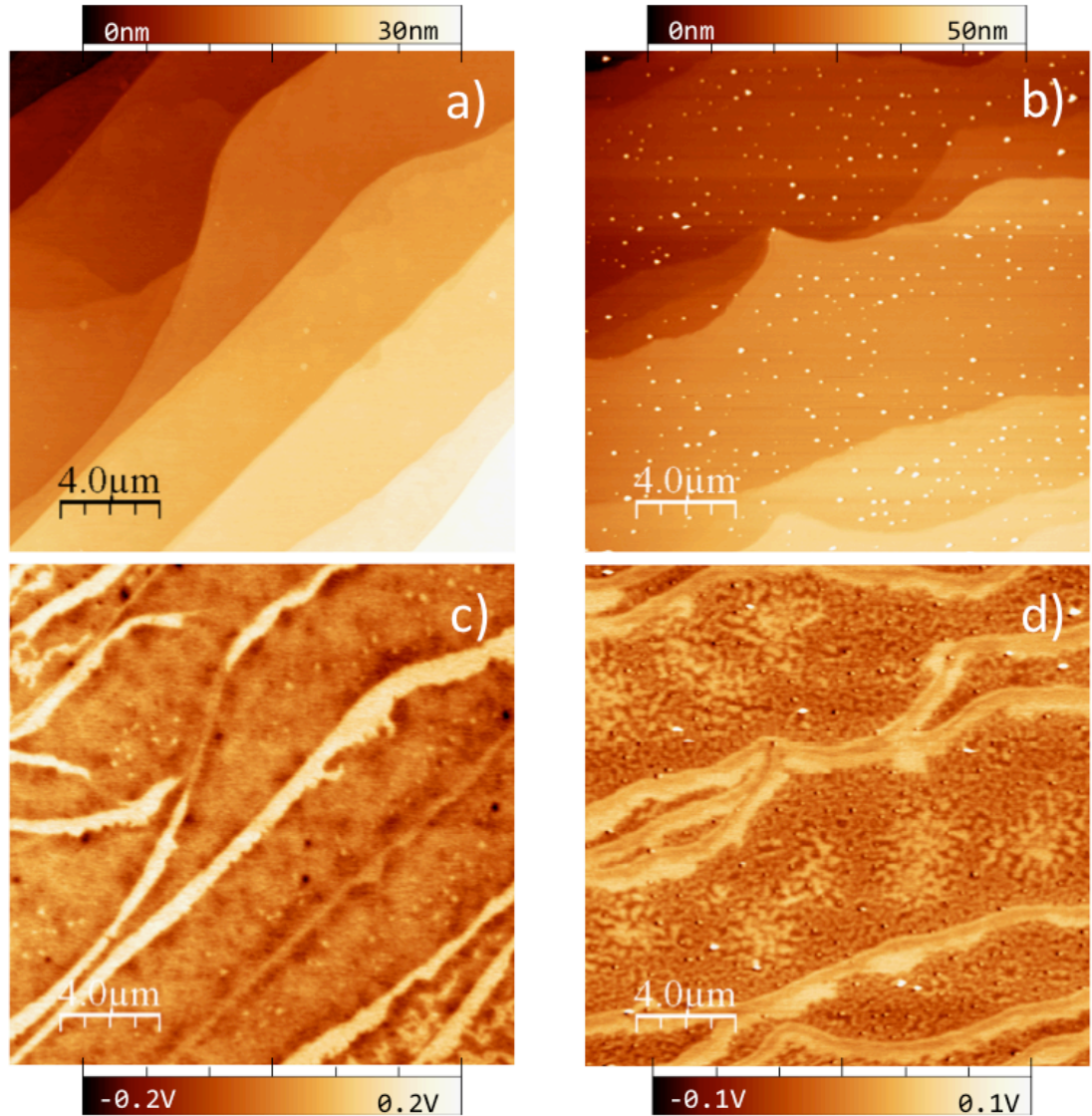

**Figure 4.** *Simultaneous topographic (a, b) and surface potential (c, d) images of SLG (left column) and QFSMG (right column) graphene on 4H-SiC(0001).*

## 4. CONCLUSIONS

In this work we have compared the structural properties of two graphene layers epitaxially grown on 4H-SiC(0001) samples. We have studied a single layer graphene formed by the CVD method and a quasi-free-standing graphene obtained by hydrogen intercalation under the CVD-grown buffer layer. The samples have been characterized extensively by several techniques that have allowed us to draw a structural and compositional model for both samples.

The results indicate that the SLG sample is formed by 73% large terraces (20-50 μm long and 5 μm wide) covered by ML graphene showing the common (6√3 x 6√3)R30º reconstruction with a high rate of coverage. In the terraces appear also 3% small islands (≈0.5μm extension) formed by BL graphene. The terraces are limited by steps with a stripe shaded carbon accumulation. There are two types of stripes. One formed by BL graphene (13% stripes≈1μm wide) wider than the other one (11% stripes≈0.5μm wide) formed by TL graphene.

On the other hand, the QFSMG sample is composed by 87% terraces (40-50 μm long and ≈15 μm wide) covered by ML graphene. However, inside the terraces some differences can be appreciated, probably due to an inhomogeneous distribution of the intercalated hydrogen. These terraces are limited by steps, again with a carbon accumulation in stripe-shape. There are two kinds of bands, those formed by BL graphene, wider (≈1 μm), with 9% coverage, and those formed by TL graphene, narrower (≈0.5 μm), with 4% coverage.

KPFM data fully agrees with the results from the LEEM/PEEM analysis, and a direct correlation between the measured images can be seen. The surface potential contrast observed arises either from thickness differences, as it is the case at the terrace edges, or from different doping levels, as we think is the case of the patches observed at the terraces of the QFSMG sample.

Lastly, the data collected from both samples indicate that near the steps there is a carbon accumulation in the form of strips, suggesting that for both kinds of samples, these regions are the nucleation centres for the formation of the graphene layer. The main difference between both graphene layers, apart from the decoupling from the substrate, is the homogeneity within the terraces. In the case of the SLG there are some islands within the terraces that represent a 3% region of the surface, while in the QFSMG case a 44% region within the terraces are covered by patches as a consequence of the hydrogenation process.

**Acknowledgements**

Financial support from EU Horizon 2020 research and innovation program under grant agreement No. 785219 (Graphene Flagship Core2-Graphene-based disruptive technologies) and Spanish MINECO (grants MAT2017-85089-C2-1-R, MAT2014-52405-C2-2-R, RYC-2014-16626) is acknowledged.